# Sequential Wnt Agonist then Antagonist Treatment Accelerates Tissue Repair and Minimizes Fibrosis

Xiao-Jun Tian[1,2,#,*], Dong Zhou[3,#], Haiyan Fu[4,#], Rong Zhang[2], Xiaojie Wang[3], Sui Huang[5], Youhua Liu[3,4,*], Jianhua Xing[1, 6, 7,*]

[1]Department of Computational and Systems Biology, School of Medicine, University of Pittsburgh, Pittsburgh, Pennsylvania, 15261, USA;
[2]School of Biological and Health Systems Engineering, Arizona State University, Tempe, AZ, 85287, USA
[3]Department of Pathology, School of Medicine, University of Pittsburgh, Pittsburigh, Pennsylvania, 15261, USA;
[4]State Key Laboratory of Organ Failure Research, National Clinical Research Center of Kidney Disease, Division of Nephrology, Nanfang Hospital, Southern Medical University, Guangzhou, 510515, China;
[5]Institute for Systems Biology, Seattle, WA, USA;
[6]Department of Physics, University of Pittsburgh, Pittsburgh, Pennsylvania, 15261, USA;
[7]UPMC-Hillman Cancer Center, University of Pittsburgh, Pittsburgh, PA 15232, USA
[#]These authors contributed equally to this work

[*]To whom correspondence should be addressed:
(Lead Contact) Jianhua Xing, Ph.D., Department of Computational and Systems Biology, University of Pittsburgh, 3501 Fifth Ave, Pittsburgh, PA 15261, E-mail: xing1@pitt.edu:
Youhua Liu, Ph.D., Department of Pathology, University of Pittsburgh, 200 Lothrop Street, Pittsburgh, PA 15261, E-mail: yhliu@pitt.edu.
Xiao-Jun Tian, Ph.D., School of Biological and Health Systems Engineering, Arizona State University, Tempe, AZ, 85287, xiaojun.tian@asu.edu:

**Abstract**

Tissue fibrosis compromises organ function and occurs as a potential long-term outcome in response to acute tissue injuries. Currently, lack of mechanistic understanding prevents effective prevention and treatment of the progression from acute injury to fibrosis. Here, we combined quantitative experimental studies with a mouse kidney injury model and a computational approach to determine how the physiological consequences are determined by the severity of ischemia injury, and to identify how to manipulate Wnt signaling to accelerate repair of ischemic tissue damage while minimizing fibrosis. The study reveals that Wnt-mediated memory of prior injury contributes to fibrosis progression, and ischemic preconditioning reduces the risk of death but increases the risk of fibrosis. Furthermore, we validated the prediction that sequential combination therapy of initial treatment with a Wnt agonist followed by treatment with a Wnt antagonist can reduce both the risk of death and fibrosis in response to acute injuries.

## INTRODUCTION

Acute injury of organs triggers inflammation and wound healing to restore tissue integrity and function. This rapid repair response involves a complex cascade of inflammatory processes that contain the damage and trigger regeneration [1,2]. During this response, instead of the signaling pathways that maintain tissue homeostasis, pathways that control self-propelling yet self-limiting program are active. During the tissue healing response, many diverse cell types, both resident and infiltrating cells, communicate through secreted signals and cell contact to modulate each other's behaviors. Some of these behaviors include proliferation and transitions between different cellular states, such as differentiated cells adopting a transient progenitor-like state or quiescent stem cells adopting an activated state [2]. A critical phase in the repair process is resolution [3], which terminates inflammation and regeneration once the damaged tissue is repaired: excess cells die, return to quiescence, or re-differentiate.

Perfect repair of injury is not always possible, especially when large areas of damage or repeated injuries occur. Consequently, the resolution process is altered resulting in the formation of scars or tissue fibrosis. In some cases, such as chronic kidney disease (CKD) [4], chronic progressive fibrosis occurs [5] due to excess proliferation of fibroblasts and deposition of extracellular matrix. Chronic progressive fibrosis impairs organ function and can ultimately lead to organ failure and death. Chronic progressive fibrosis is not simply "incomplete" or "altered" repair; instead, it results from excessive repair activity and failure of resolution and is, thus, a maladaptive response. Fibrosis impairs organ function in heart, kidney, and liver diseases, and is a complication that impairs function after organ transplantation and organ health following surgery. For example, ischemic kidney injury is a common consequence of cardiac surgery [6].

Here, we used a mouse model of ischemic kidney injury to explore acute injury repair and chronic fibrosis. Exposure to toxins or hypoxic stress can lead to loss of epithelial cells in the renal tubules, causing either acute kidney injury (AKI) with kidney failure or CKD that progresses irreversibly to end-stage renal insufficiency. AKI and CKD are global health challenges with limited treatment options [7]. These two conditions are often mechanistically and clinically linked [8-10]. Although some patients completely recover from AKI, others progress to CKD [11,12]. The determining factors for the transition from AKI to CKD are not clear. The risk of CKD increases if a patient survives a single episode of AKI and further increases with repetitive AKI episodes [13,14]. In mice, exposing the kidney to mild ischemia prior to injury (ischemic preconditioning) protects against renal damage from a subsequent AKI [15-17]. Clinical trials of ischemic preconditioning for AKI associated with cardiac surgery are ongoing [18,19].

Mechanistically, kidney tissue injury stimulates two connected cellular networks (tubular cells and fibroblasts) through signals mediated by the Sonic hedgehog (Shh) and Wnt pathways. Shh released by the kidney tubular cells stimulates proliferation of activated fibroblasts to promote tissue repair [20]. Wnt produced by both tubular cells and fibroblasts participates in both repair of injured tissue and stimulation of fibrosis [21-24]. In particular, Wnt signaling stimulates cell state transitions, including epithelial-to-mesenchymal transition (EMT) and partial EMT (pEMT), which contribute to both tissue repair and many chronic fibrotic diseases [4,25,26]. After AKI, some kidney tubular epithelial cells undergo pEMT and some activated fibroblasts become myofibroblasts, the latter of which is a critical event in the development of fibrosis [27,28]. The Wnt-stimulated process of myofibroblast expansion could serve as a drug target for limiting fibrosis and development of CKD. Consistent with this hypothesis, excessive and prolonged activation of Wnt signaling promotes myofibroblast proliferation and fibrogenesis and correlates with progression to CKD [23,24].

The apparent seemingly contradiction of the beneficial effects of ischemic preconditioning and the established detrimental role of Wnt activation on fibrosis development has led to a long-standing puzzle in the field of AKI and CKD studies. To resolve this controversy, here we combined quantitative experimental studies with a mouse kidney injury model and a nonlinear dynamical systems model of the cellular interaction network composed of a tubular epithelial cell module and a fibroblast cell module, involving three cell states in each module, and regulatory signals mediated by Shh and Wnt. We used this cellular network model to investigate the transition from acute injury and repair to chronic tissue damage and fibrosis. The model provided a mechanism for the multiple outcomes of kidney injury found experimentally in a mouse model. Furthermore, the model predicted a potential increased risk of fibrosis of ischemic preconditioning despite its well-known protective effect, which were validated in a mouse kidney ischemia system. Importantly, we developed a Wnt pathway-targeted treatment regimen involving sequential agonist then antagonist treatment to accelerate tissue repair and prevent fibrosis with the guidance of computational multi-objective optimization. These findings are particularly relevant in the context of surgery when the timing of the ischemic episode is known and treatments can be initiated and strategically planned to optimize positive outcomes.

**RESULTS**

**Mouse models reveal four discrete terminal states depending on the duration of acute kidney injuries**

We established a mouse model for AKI [the ischemia-reperfusion injury (IRI) model] to study whether varying a single input variable (the severity of transient ischemia) produces qualitatively distinct repair outcomes. If so, then the system is a complex non-linear dynamical system, which has the property that multiple distinct stable outcomes arise in response to a continuous range of a perturbation. We tuned the severity of kidney injury by varying the duration of ischemia [23]: mild (5 min), moderate (10, 15, and 20 min) and severe (30 min) IRI. All mice exposed to mild IRI survived, three out of seven mice exposed to each of the moderate IRI conditions survived, and only 10 out of 16 mice exposed to severe IRI survived. Surviving mice recovered for 30 days; then we analyzed the kidney tissue by Periodic Acid-Schiff (PAS) (Fig. S1) to evaluate the severity of injury based on morphological changes and marker expressions. We stained for Wnt1, Vimentin, fibroblast specific protein 1 (FSP-1), α-smooth muscle actin (α-SMA), and platelet-derived growth factor receptor-β (PDGFR-β) by immunohistochemistry to monitor activation of fibroblasts and pEMT (Fig. 1A, B). For animals subjected to mild IRI, we observed that all markers returned to basal levels like those in control mice not exposed to IRI. The response to 10-min IRI was more variable: Some mice had a small but statistically significant increase in Wnt1, FSP-1, and vimentin, but not PDGFR-β; only one mouse had a statistically significant increase in α-SMA. The other three experimental groups displayed a significant increase in all five markers. Mice exposed to the severe IRI displayed the highest amounts of all of these markers and had large fibrotic patches (Fig. 1A, Fig. S1). In contrast, small and scattered fibrotic patches were observed in mice after 15- or 20-min IRI (Fig. S1).

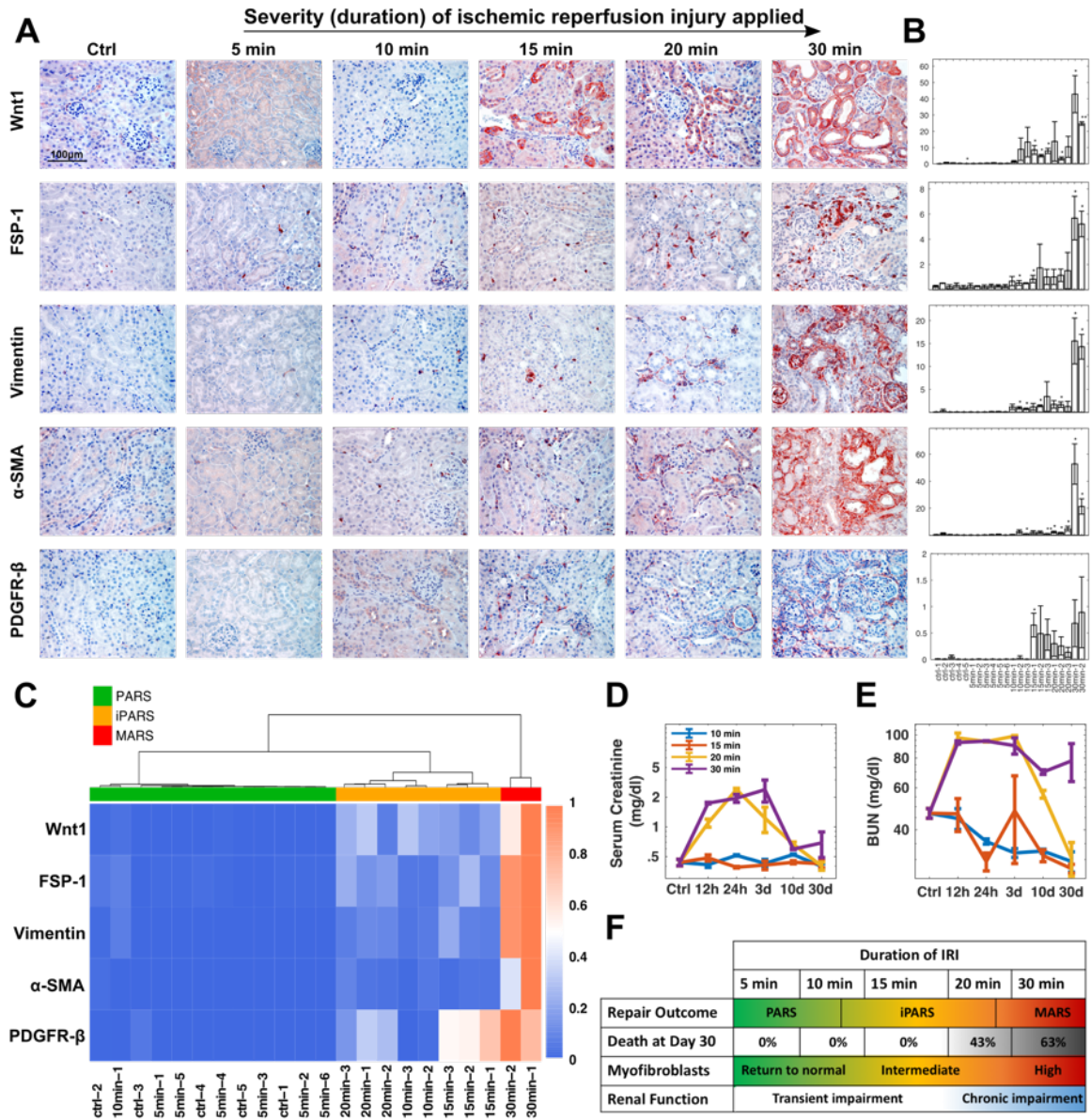

***Figure 1.*** *Mouse models show existence of four possible outcomes depending on the duration of acute kidney injuries. (A) Representative micrographs showing expression of Wnt1, FSP-1, Vimentin, α-SMA, and PDGFR-β in control and diseased kidneys 30 days after varying degrees of IRI. Scale bar, 100 µm. (B) Quantification of marker levels. Each column indicates an individual mouse. Data is represented as mean ± s.e.m of 3 slices for each marker each mouse. (C) Hierarchical clustering analysis of the marker levels on day 30 after different duration of IRI treatment. (D-E) Temporal profiles of renal function markers, serum creatinine (D) and BUN (E) after various duration of IRI. n=3, Data is represented as mean ± s.e.m. (F) The fraction of the dead mice, repair outcomes, myofibroblast expansion and renal function depends on the duration of IRI.*

Cluster analysis with these markers 30 days after injury divided the surviving mice into three readily distinguishable groups (Fig. 1C) representing the treatment groups, indicating that these markers adequately captured the differences between the mild, moderate, and severe injury conditions. In the mice in the 5-min IRI group, the markers returned to basal, the tissue was morphologically indistinguishable from the control mice, and the mice clustered with the control mice, indicating that the repair response was complete and resolved. Hence, we defined the kidney tissue state in the mice of this cluster as a "perfect adaptive response state" (PARS). The markers of mice in the 15- and 20-min IRI groups indicated the persistence of low levels of Wnt and low numbers of residual activated fibroblasts and pEMT cells, thus we defined this as an "imperfect adaptive response state" (iPARS). The third cluster of the 30-min IRI group had kidneys with high amounts of Wnt and markers of myofibroblasts and pEMT cells, thus, we defined this as a "maladaptive response state" (MARS).

To determine how increasing severity of IRI affected kidney function, we evaluated kidney function based on serum creatinine and blood urea nitrogen (BUN), which are routinely used to assess kidney function in patients. By 30 days after IRI, only the mice in the severe IRI group exhibited compromised kidney function (Fig. 1D-E). Consistent with the severe IRI and 20-min moderate IRI group exhibiting the most impairment in kidney function (highest serum creatinine and BUN), those two groups of mice had several animals die in the first two weeks after the IRI (Fig. 1F).

These data indicated that the kidney response to IRI is a complex non-linear dynamical system. A single quantitative variable, the duration of ischemia, produced four qualitatively distinct outcomes: (i) full pathological recovery (PARS), (ii) imperfect recovery with mild lasting tissue changes but restored kidney function (iPARS), (iii) a maladaptive response (MARS) with signs of fibrosis and renal dysfunction, and (iv) death.

**A mathematical model of the regulatory network of the kidney response to injury accurately recapitulates the four observed outcomes**

We first asked whether the four distinct and robustly separable outcomes in the animals, triggered by varying the magnitude of one experimentally controllable parameter, reflect inherent dynamical behaviors of the cell-cell interaction network that orchestrates the repair response. We constructed a canonical model with the key components involved in kidney repair (Fig. 2A, Table S1-2 ). We represented the model with ordinary differential equations with eight variables representing the rates of change in the abundance of cells and the two signaling mediators (Shh, Wnt) (SUPPLEMENTARY MATERIALS). The cellular components of the network are represented by two modules: the tubular module (representing the tubular epithelial cells) and the fibroblast module, each containing three distinct cell states. Shh and Wnt produced by the cells in the cellular modules mediate the interactions between the two modules. The cells proliferate, undergo state transitions, or cell death under the influence of these two signaling molecules in response to ischemic injury ('insult' in the model). The outcome variables include (i) the fraction of dead tubular epithelial cells (Fig. S2, panel A, see SUPPLEMENTARY MATERIALS: The fate of organism death in the mathematical model), and (ii) the abundance of myofibroblasts (Fig. S2 ). We selected the number of myofibroblasts as a determinant of the outcome related to fibrosis, because the abundance of myofibroblasts is a major determinant of this pathology [29]. Using these variables, we set to examine whether the system could reproduce multiple clinically and experimentally observed outcomes, such as full recovery without fibrosis (no dead tubular epithelial cells and no increase in myofibroblasts compared with the uninjured state), fibrosis (indicated by the steady state abundance of myofibroblasts), or organismal death (modeled as the number of dead tubular epithelial cells exceeding a preset threshold).

Under physiological conditions, the number of fibroblasts residing in the interstitial compartment is low [27]. We model these as 'resident fibroblasts' (Fig. 2A). In response to insult, some healthy epithelial cells ('tubular epithelium' in the model) enter the injured state ('injured tubular cells') and promote self-renewal of neighboring tubular epithelial cells. In the model, Shh promotes resident fibroblasts to transition into the state 'activated fibroblast' in which they proliferate and from which they can further transition into the state 'myofibroblast' [20,30]. The myofibroblasts secrete Wnt, which stimulates proliferation of both activated fibroblasts and myofibroblasts, establishing a positive feedback loop between the myofibroblasts and Wnt production.

Wnt also affects the tubular module [31], further promoting self-renewal and repair of injured tubular epithelial cells). Wnt also stimulates tubular epithelial cells to undergo a state transition to pEMT ('partial EMT tubular cell') [24,27,28,32,33], which also secretes Wnt [28]. This establishes a second positive feedback loop between the partial EMT tubular cell and Wnt production.

To determine the outcomes produced by the model, we set the parameters to those for mice (see Table S1-2 for values of parameters and initial states, see SUPPLEMENTARY MATERIALS: Parameter Estimation and Justification), then monitored the distinct stable steady states produced by the model as a function of myofibroblast number by either applying a range of insult intensities (Fig. 2B) or starting from a range of initial conditions (Fig. S2B). The simulated time course of the number of myofibroblasts showed that the network existed in one of three stable steady states with respect to myofibroblast abundance: return to baseline, persistent intermediate numbers of myofibroblasts, or persistent high numbers of myofibroblasts (Fig. 2B, Fig. S2B). Increasing insult duration and incorporating the number of dead tubular epithelial cells into the outcome of the model resulted in the separation of organismal death from states associated with survival, and separated each of the three survival-associated steady states into those representing severe fibrosis associated with high numbers of myofibroblasts (representing MARS in the mouse experiments), low numbers of residual myofibroblasts (representing iPARS), and no fibrosis, no increase in myofibroblasts from baseline, and complete recovery of tubular epithelial cells (representing PARS) (Fig. 2C). Further analysis of the cell network (Fig. S2C) suggested that the three steady states associated with survival are stable with respect to other system variables.

Consequently, we could parameterize insult by duration and strength in a 2D-phase diagram, which divided into the four outcome domains and revealed that a short and severe insult had a similar effect as a longer but less severe one (Fig. 2D). To account for the stochastic variability between mice, we simulated an ensemble of models with parameters randomly and uniformly chosen within ± 20% of its default value and measured the fraction of mice exhibiting each of the four outcomes. By representing the data on a 2D-phase diagram, we showed that the fraction of mice displaying PARS decreased with increasing duration of the insult and that the fraction of mice suffering from the next higher-level severity of disease (or death) increased with increasing duration of the insult (Fig. 2E). This outcome is consistent with the experimental results observed for the mice (Fig. 1F).

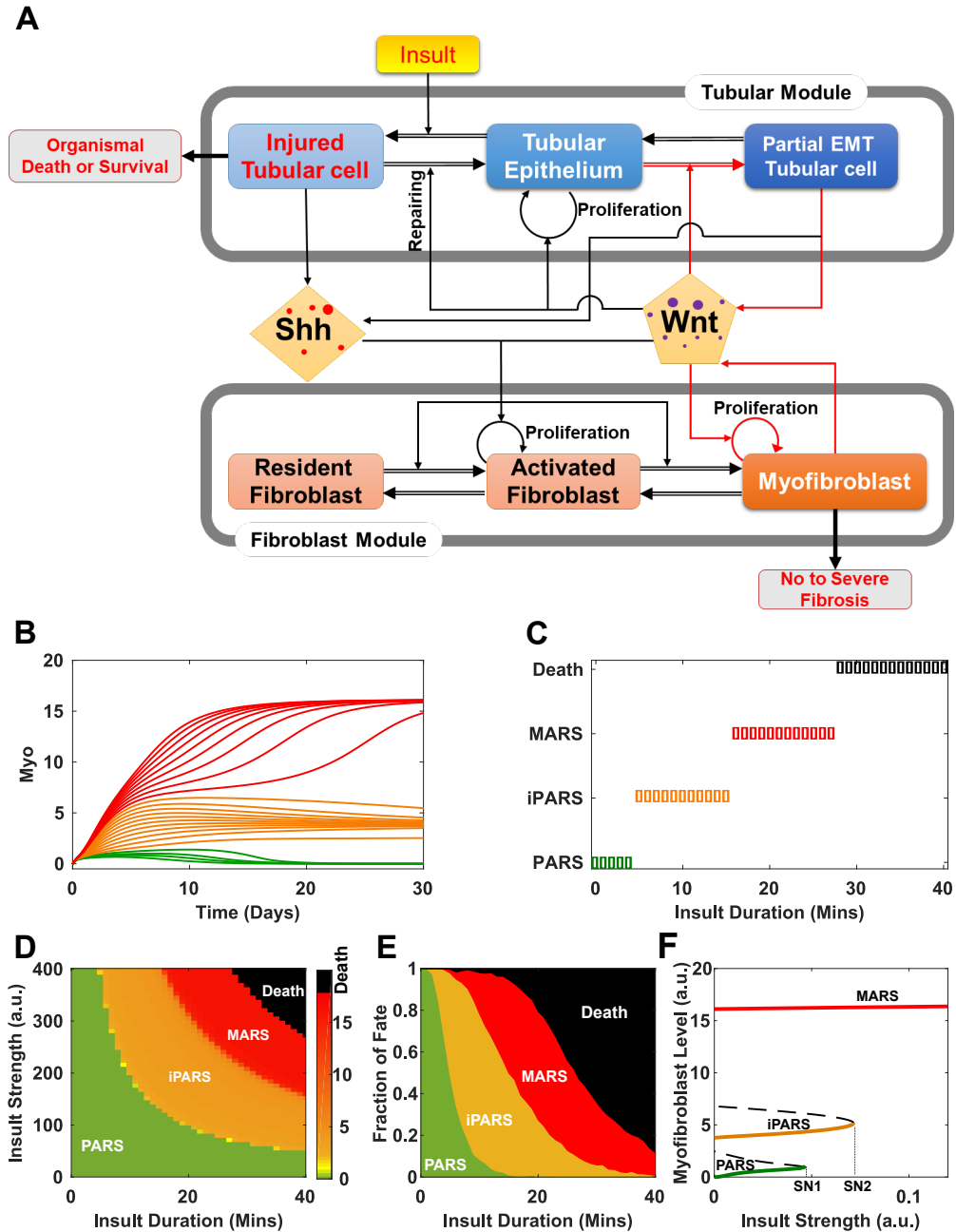

***Figure 2*** *Theoretical analyses reveal the mechanism of the four outcomes in response to renal injury. (A) Cell-cell communication model for renal homeostasis and fibrosis. The positive feedback loops are highlighted in red. (B) Simulated time series of the number of myofibroblasts under different durations of insult. (C) Repair outcomes under insult of fixed strength but increasing durations. (D) Phase diagram of repair outcomes (indicated by myofibroblast level) in the space of insult strength and duration. (E) Fraction of outcomes as a function of insult duration sampled from 1000 independent simulations. (F) Bifurcation diagram of the myofibroblast level with respect to insult strength shows three possible survival outcomes depending on the strength of the insult.*

**Bifurcation analysis reveals increasing insult severity triggers distinct irreversible states**

Because the three outcomes —PARS, iPARS and MARS— map to stable steady states of the cellular network model (Fig. 2A), we performed an analysis of the irreversibility of the states using a one-parameter bifurcation analysis. Under mild insult, the system ended in PARS (Fig. 2F, green line) and returned to the state prior to the insult after removing the insult (with zero myofibroblasts at zero insult strength). However, increasing insult severity above a threshold (Fig. 2F, SN1) resulted in the system crossing a first bifurcation point. Consequently, the system "jumped" to iPARS (Fig. 2F, orange line). In this state, the system exhibited hysteresis, because reducing the insult strength to zero did not restore a completely healthy tissue composition. Instead, low levels of myofibroblasts remained. This is consistent with the preservation of kidney function that we observed in the mice (Fig. 1 D-E). With further increase of insult strength beyond a second critical point (Fig. 2F, SN2), the system jumped to MARS (Fig. 2F, red line), characterized by high abundance of myofibroblasts that persisted even after the insult was reduced to zero. Thus, the model predicted that both iPARS and MARS are irreversible states (for full bifurcation diagram, see Fig. S2D). Parameter sensitivity analysis (Fig. S2E) revealed that all four outcomes existed within a parameter range of 15% increase or decrease of most parameters, indicating that existence of these outcomes is insensitive to heterogeneity of individual mice or our model parameter choice.

**Preconditioning to iPARS reduces risk of death but increases risk of fibrosis**

Mild ischemic insults have a protective role against subsequent insults in several organ systems, such as brain, heart, liver, and kidney. Protective preconditioning is being explored as a therapeutic modality for kidney injury associated with cardiac surgery [15-17]. We hypothesized that mice surviving a previous AKI would be in iPARS, which would protect them against subsequent ischemic events but would increase their risk of CKD after subsequent ischemic events. We examined whether protective preconditioning is a property of our kidney cellular interaction network. We simulated a moderate renal insult (10 min) to place the system into iPARS, characterized by the absence of dead tubular cells and the presence of residual myofibroblasts (Fig. 3A, solid black lines). Following a second moderate insult (15 min), the percentage of dead tubular cells reduced faster, indicating faster repair and regeneration, in the preconditioned system (Fig. 3A left, red solid line) than in the control system without preconditioning (Fig. 3A left, black dashed lines). Thus, our system exhibited properties of protective preconditioning.

However, the preconditioned system produced a high level of myofibroblasts (Fig. 3A right, red solid line). Thus, the model predicted that ischemic preconditioning increases the risk of CKD. Because the preconditioned state represents iPARS, Wnt, myofibroblasts, and pEMT tubular cells are predicted to persist. This prediction is consistent with clinical observations that some patients, who survive an episode of AKI with no obvious residual impairment of renal function, have significantly increased risk of developing CKD [13,14]. Using our computational model, we also tested if repetitive mild AKI, with episodes of different duration and different spacing between episodes, increased the risk of CKD. Our model indicated that some dead tubular cells persisted after each episode and that the level of myoblasts increased after each episode (Fig. S3A-B). Additionally, prolonged and more frequent exposure to mild AKI resulted in quicker development of CKD (Fig. S3C) Thus, our simulations recapitulated both the clinical and animal model observations that repetitive mild AKI also increased the risk of CKD [34,35].

The previous simulations tested the induction of iPARS on the outcome of a mild second insult (15 min). We also tested the induction of iPARS on the outcome of a stronger second insult (30 min). An insult of this intensity is predicted to sometimes lead to death in the naïve control systems (Fig. 2C, E). Our model predicted that iPARS reduced the peak percentage and cumulative sum of dead tubular cells (Fig. 3B, black dashed lines), thus preventing organism death (Fig. 3B right, dotted vertical line). This reduction in dead tubular cells was predicted for a large range of durations of the second insult (Fig. 3C), indicating that iPARS has a protective role in reducing the risk of death from subsequent severe renal injury.

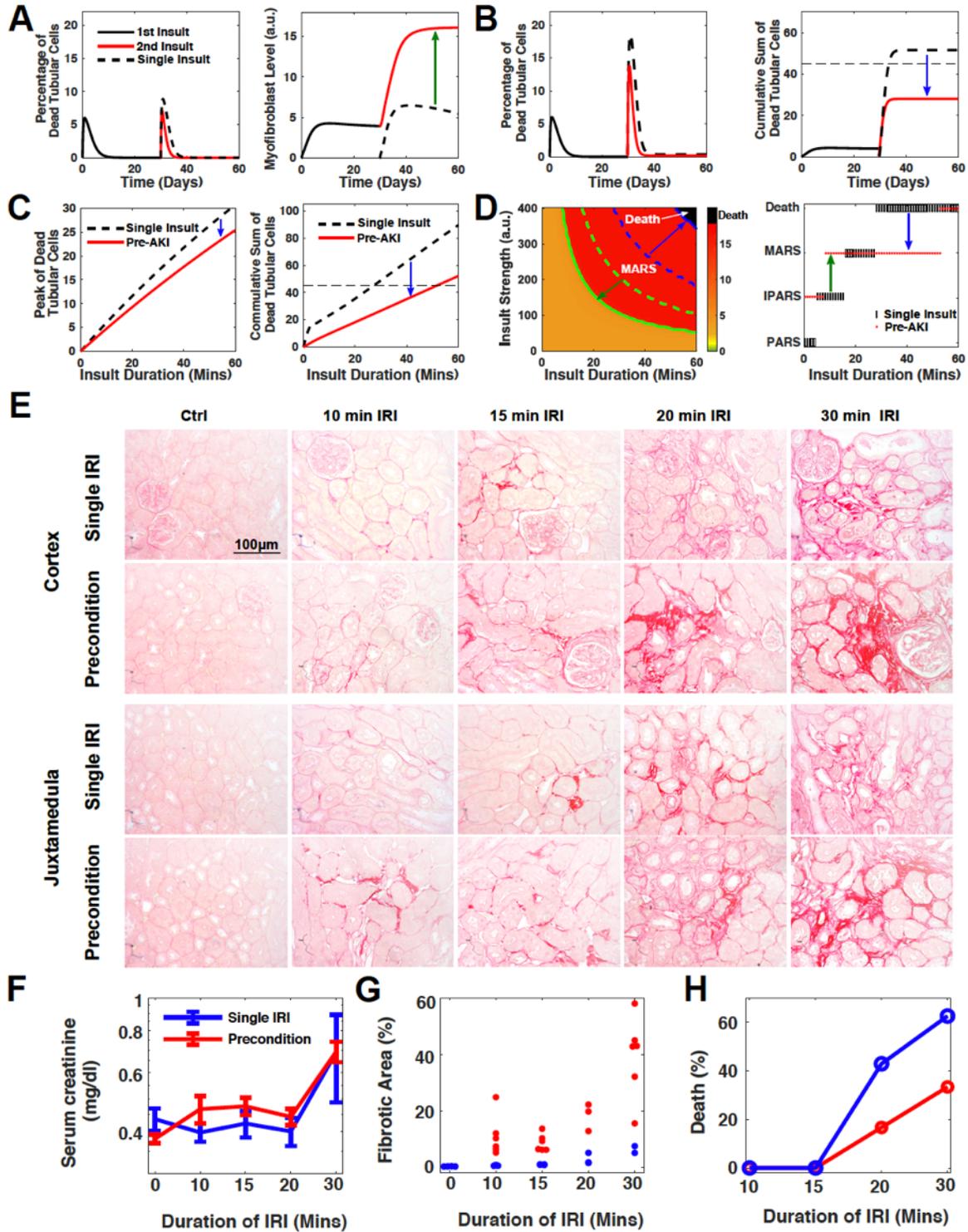

*Figure 3. Double-edged sword effects of pre-fibrotic iPARS, reducing the risk of death but increasing the risk of MARS. (A) Temporal profiles of the percentage of dead tubular cells and the myofibroblast level under one 10 min insult (dashed lines) and two successive insults with one second 15-min insult occurring after 30 days after the first (solid lines). (B) Same as panels A-B except with a 30-min second insult. (C) Peak level of dead tubular cells and maximum level of the cumulative death tubular cells as a function of insult duration. The system is either preconditioned with a first moderate insult (solid red lines), or without preconditioning (black dash lines) as control. (D) (Left) Phase diagram of repair outcomes (indicated by myofibroblast level) in the space of insult strength and duration. Similar as Fig. 1D, except preconditioned with a moderate insult (duration = 10 mins, 30 days of recovery after first insult). Green lines indicate the threshold of MARS with (solid) and without (dashed) preconditioning. Blue lines indicate the threshold of death with (solid) and without (dashed) preconditioning. (Right) Repair outcomes under insults of fixed strength but increased durations with (red dots) and without (black boxes) preconditioning (duration = 10 mins, 30 days of recovery). (E) Representative micrographs of Picro Sirius Red staining show collagen deposition in kidney cortex and juxta medulla area in both group of pre-conditioned mice and single-IRI mice 30 days after varying degree IRI. Scale bar, 100 μm. (F-H) Dependence of the level of serum creatinine (F), the fraction of the fibrotic area (G), and the animal death fraction (H) on the duration of IRI 30 days after IRI. Preconditioning: 10 min IRI followed by 30 days of recovery.*

Taken together, our model predicted that the relative benefit of ischemic preconditioning of kidneys to place them in iPARS depended on the severity of the second insult. When the second insult was severe, the benefit of ischemic preconditioning in reducing the risk of death exceeds that of the increased fibrotic risk associated with MARS, which reflects an increased risk of CKD. Ischemic preconditioning raised the threshold of death and lowered that of MARS (Fig. 3D left, blue and green arrows), thus expanding the insult conditions (parameters in the model) associated with survival at the cost of increasing the risk of chronic fibrosis (Fig. 3D right).

To test the mathematical predictions, we preconditioned mice by subjecting them to 10-min IRI. After 30 days of recovery, the mice were exposed to various durations of a second IRI from 10 min to 30 min. We observed larger fibrotic patches in both cortex and medulla in all groups of preconditioned mice subjected to a second IRI compared to the animals only subjected to a single IRI (Fig. 3E). Serum creatinine levels did not reveal an obvious change in renal function between mice subjected to preconditioning and a second IRI and those subjected to a single IRI (Fig. 3F). However, quantification of the fibrotic areas revealed a significant increase of fibrosis in the preconditioned mice compared to that of the single IRI group (Fig. 3G), consistent with the predictions of the model. Also consistent with the model predictions, we observed significant protection from death in the animals subjected to preconditioning, especially for cases of more intense insult (Fig. 3H). Thus, these animal experiments recapitulated the opposing effects of preconditioning predicted by the model: Preconditioning reduced death rate at the cost of increased fibrosis.

**Wnt-mediated positive feedback loops modulate the repair response and risk of progression to CKD**

To explore the molecular origin of the four AKI outcomes, we focused on the positive feedback loop formed by the pEMT tubular cells and Wnt (Fig. 2A). We selected this feedback loop because it represents a critical aspect of tissue repair (pEMT tubular cells) and a major contributing factor to fibrosis (Wnt secretion that stimulates myofibroblast expansion). With the model, we tested the effect of altering Wnt signaling on the outcome of injury (Fig. 4, Fig. S4).

From a clinical standpoint, Wnt signaling represents a potentially pharmacologically manipulatable aspect of the injury response, thus we focus on the results of those simulations.

We simulated the outcomes in the cellular network model by perturbing the activity of Wnt. When we reduced Wnt in the network model, the system tolerated higher insult intensities before jumping to either iPARS or MARS (Fig. 4A, blue lines compared to black lines). When we increased Wnt, the system jumped to iPARS and MARS at lower insult intensities (Fig. 4A, compare red lines to black lines). These results indicated that reducing Wnt in the presence of a continuous low-intensity insult was protective, increasing the threshold of MARS and making the maladaptive response less likely; whereas increasing Wnt had the opposite effect under these conditions.

We also simulated the outcomes in response to a single transient insult of varying duration representing AKI, while increasing or decreasing Wnt (Fig. 4B-E). When Wnt was reduced, there were only three states: PARS, iPARS, and death. The insult durations that resulted in death or iPARS increased, expanding those states and eliminating MARS (Fig. 4B-C). Conversely, when Wnt was increased, the insult durations that resulted in MARS increased and fewer resulted in death or iPARS (Fig. 4D-E), thus raising the threshold for death and lowering that for MARS.

We simulated the effect of Wnt agonists added prior to a single transient insult of varying duration. The model predicted that pretreatment with Wnt agonists raised the threshold of death and expanded the insults that resulted in MARS (Fig. 4F). The predicted reduction in death is consistent with a report that treatment with Wnt reduced kidney damage and improved renal function in an IRI rat model [36]. Inspection of the transition between iPARS and MARS (Fig. 4F) suggested that the effect of Wnt pathway activity was biphasic: The threshold of MARS first increased (reflecting a safer effect of Wnt treatment) and then decreased as Wnt agonists increased. These results predicted that the therapeutic window for Wnt activation that minimizes both the risk of death from acute injury and the development of chronic fibrosis is narrow.

Consistently, increasing the Wnt secretion rate in the model increased the death threshold and decreased the MARS threshold (Fig. 4F). That is, the renal system faces a fundamental trade-off between CKD (represented as MARS) and death, so reducing the risk of death increases the risk of CKD and vice versa. Thus, the model predicted that the consequences of manipulating Wnt consistently throughout the injury response are complicated and depend on the condition (low intensity and continuous injury, like CKD, or transient injury, like AKI). In CKD, reducing Wnt is predicted to be protective. However, in AKI, reducing Wnt is predicted to increase the risk of death. Furthermore, adding Wnt agonists as a pretreatment strategy is predicted to have a narrow optimal activity window.

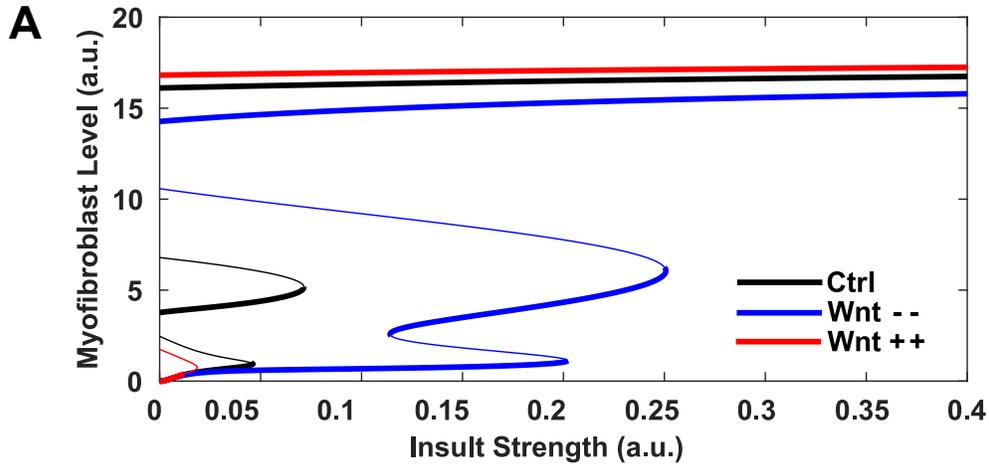
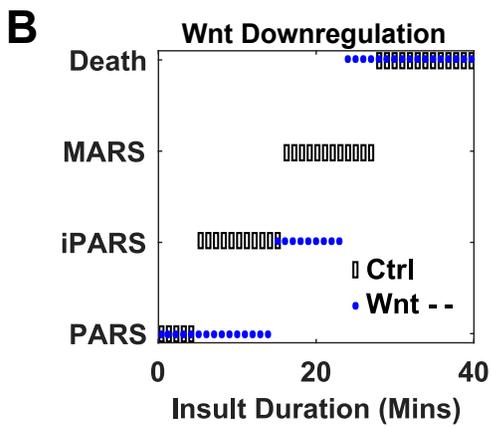
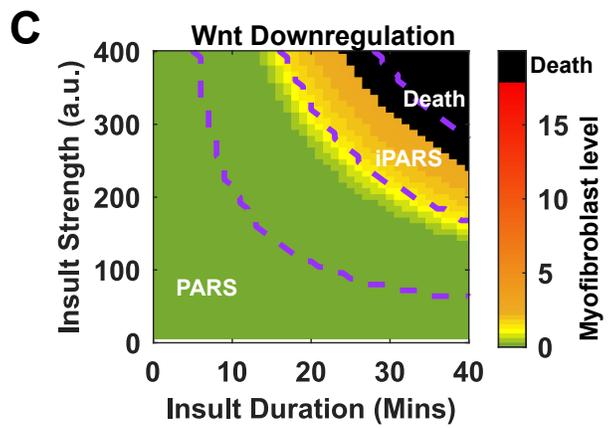
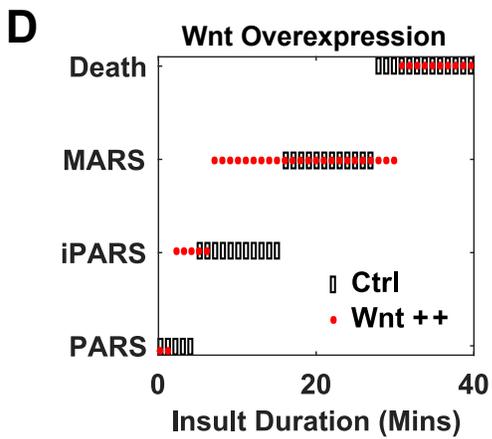
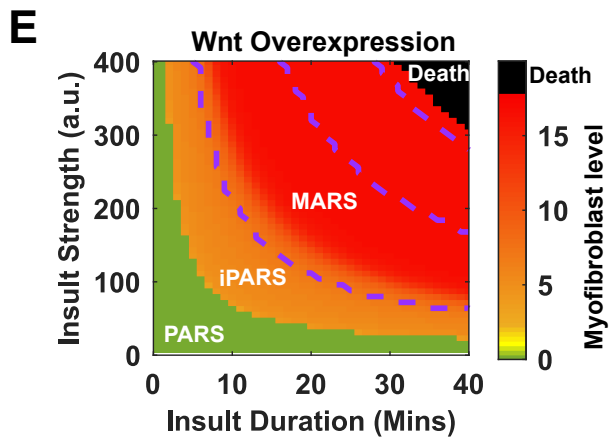
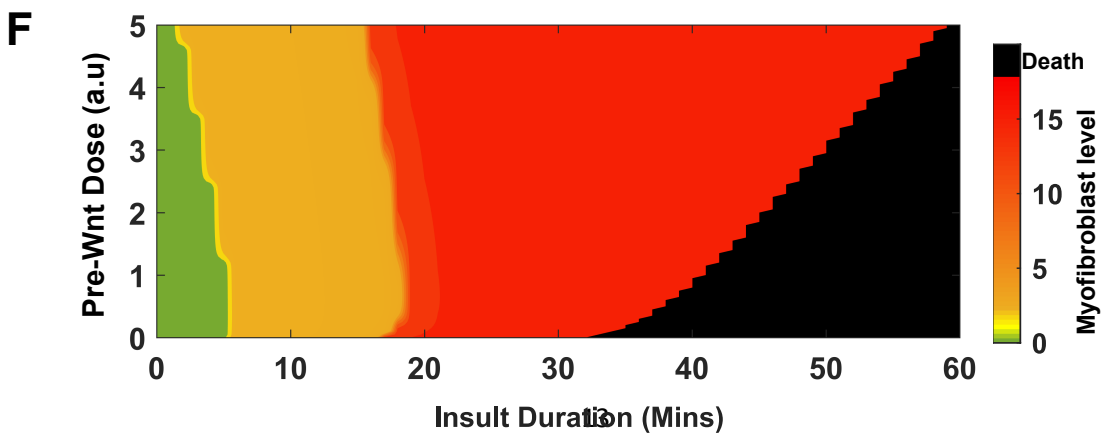

*Figure 4.* The pEMT-Wnt axis modulates repair response and progression risk to CKD. (A) Bifurcation diagram with overexpressed Wnt (red lines) or downregulated Wnt (blue lines) in contrast to the control (Ctrl, black lines). (B) Repair outcomes under insults of fixed strength but increasing durations with (blue dots) and without (black boxes, control) Wnt downregulation. (C) Phase diagram of repair outcomes (indicated by myofibroblast level) in the space of insult strength and duration with downregulated Wnt. Dashed lines indicate boundaries in the control system (without Wnt downregulation). (D) Repair outcomes under insults of fixed strength but increasing durations of IRI with (red dots) and without (black boxes, control) Wnt overexpression. (E) Same as panel C except with Wnt overexpression. Dash lines represent the boundaries in the control system (without Wnt overexpression). (F) Effect of Wnt agonist treatment before AKI on repair outcome under various insult duration.

**Achieving optimal treatment of AKI by dual targeting of repair and resolution dynamics**

In view of the trade-off between repair and fibrosis and the narrow therapeutic window of Wnt activity predicted by our cellular interaction model, we hypothesized that an optimal strategy targeting Wnt would combine activation and inhibition in a temporally controlled manner, thereby promoting both the repair and resolution phases of the injury response. Thus, we formulated it as a computational multi-objective optimization problem and used a Metropolis search to evaluate Wnt-targeted regimens by varying the schedules, durations, and dosages of both Wnt antagonists and agonists (Fig. 5A, see Method). The ideal strategy minimizes both the risks of death and fibrosis and achieves the fastest recovery using the lowest doses of drugs with shortest treatment time. Practically the optimization process is to minimize a score function that includes all these factors with the highest weight given to reducing the risk of death, followed by the weight for reducing the risk of fibrosis. As expected, the optimal strategy required treatment with a Wnt agonist at the beginning followed by treatment with a Wnt antagonist (Fig. 5B, Fig. S4A). Simulation of 1,000 mice, each with parameters drawn from a distribution to account for individual heterogeneity confirmed the robustness of the treatment scheme with the most mice under a combined treatment of Wnt agonist and antagonist reaching PARS and a small fraction reaching iPARS (Fig. 5C). Our model also predicted that the optimal treatment design depends on insult duration (Fig. S4B-C).

To validate our therapeutic strategy, we used a 30-min IRI and tested four different treatment designs in mice: IRI only, a single Wnt1 pretreatment prior to IRI, Wnt signal inhibitor ICG-001 treatment from administered every day from day 4 to day 13 after IRI, and combined treatment with Wnt1 before IRI and ICG-001 after IRI. Histochemical analysis showed that fibrosis was markedly reduced in the groups receiving ICG-001 alone or in combination with Wnt1 pretreatment; pretreatment with Wnt1 alone increased the severity of fibrosis (Fig. 5E-F). Compared to the IRI only group, we observed a reduction of components of the extracellular matrix, indicating reduce extracellular matrix deposition, in the groups with combined treatment or ICG-001-only treatment, whereas Wnt1 pretreatment alone increased extracellular matrix deposition (Fig. 5G-H). Either Wnt1 pretreatment alone or the combination treatment enhanced survival to this severe insult (Fig. 5I). In the absence of ischemic insults, administration of either Wnt1 or ICG-001 alone had little effect (Fig. S5), recapitulating our previous findings [23,37].

Taken together, the prediction from theoretical analysis and preclinical studies in a mouse model for AKI and CKD suggested that combined sequential treatment with Wnt agonists and antagonists can overcome the dilemma of the double-edged-sword effect of Wnt pathway activity and minimize both animal death and kidney fibrosis.

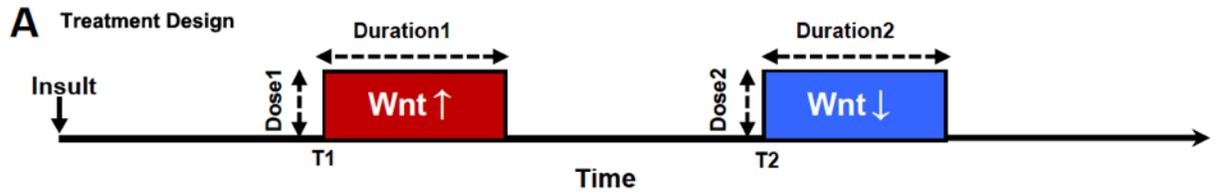
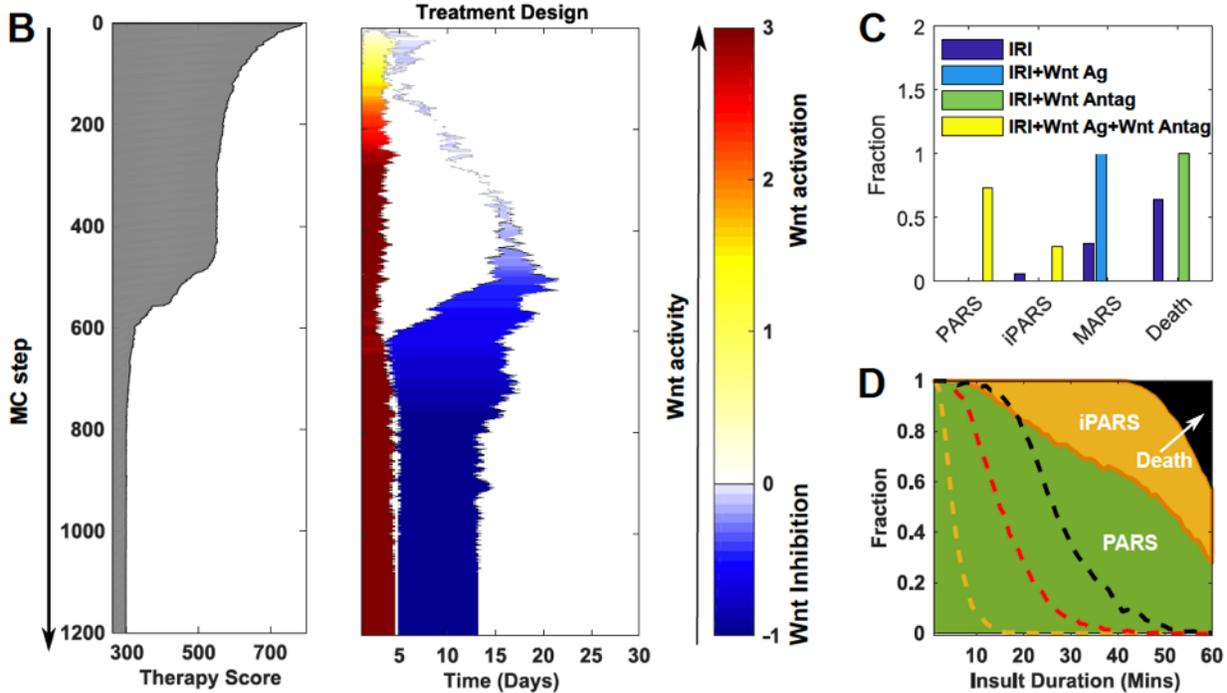
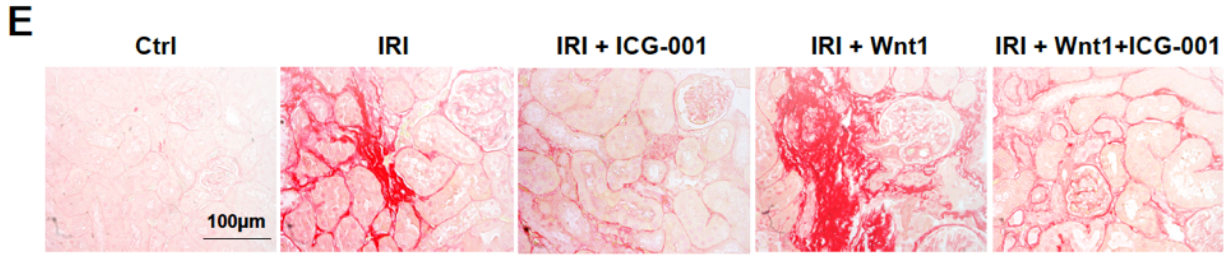
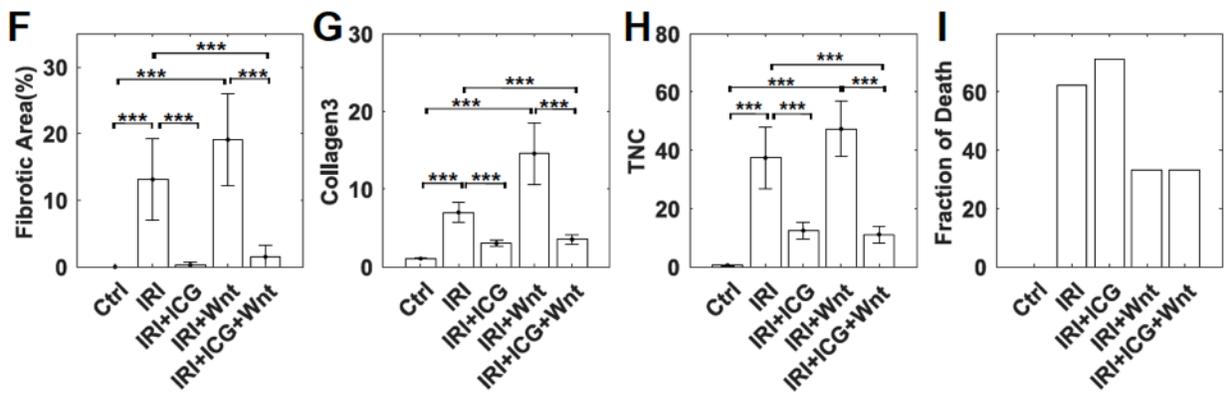

*Figure 5. Temporal regulation of Wnt represents the optimal dynamic treatment design for AKI through temporal regulation of Wnt. (A) Schematic of treatment design combining Wnt and its inhibitor, with parameters to be optimized including the doses, timings, and durations of Wnt and inhibitor. (B) Representative searching trajectory of the therapy score function and treatment design over Monte Carlo steps. The weights of different factors [λ1, λ2, λ3, λ4, λ5] in the score function are set as [500, 10, 1, 1, 1]. (C) Fraction of various repair outcomes under different treatment designs - Wnt agonist only (Wnt Ag), Wnt antagonist (Wnt Antag), or combination of Wnt agonist and antagonist - and the control system without treatment, each sampled from 1000 simulations. (D) Fractions of repair outcomes as a function of insult duration with the optimal design for the 30-min insult. Dashed lines are corresponding region boundaries without treatment (yellow: PARS/iPARS, red: iPARS/MARS, black: MARS/Death). Each result was sampled from 1000 independent simulations. (E-I) Representative micrographs of Picro Sirius Red staining(E), the fraction of fibrotic area (F), quantitative real-time RT-PCR (qRT-PCR) for collagen type III and TNC (G-H) on day 30 and the fraction of animal death (I) (within 30 days) after 30-min IRI with pre-administration of Wnt treatment or with Wnt inhibitor ICG-001 administration from day 4 to day 30 or combination of Wnt pre-administration and ICG-001 administration from day 4 to day 30 or without any treatment. Scale bar, 100 μm. \*\*\*P < 0.01 versus labeled groups(n=4-6). For panel I, the total number of the mice is 9-16 for each group. For F-H, data are represented as mean ± s.e.m.*

## DISCUSSION

Tissue injury and its repair pose a conundrum that lies at the intersection between recovery from acute disease and development of chronic disease. The unleashing of a well-orchestrated and rapid response that fixes the parenchymal tissue defect through a transient mesenchymal proliferative response must be tightly controlled to prevent chronic fibrosis caused by the response itself. Thus, tissue repair that ensures survival ranges from full reconstitution to chronic disease. The latter represents an evolutionary trade-off that prioritizes survival over avoidance of proliferative fibrotic disease.

We used an established animal model for acute ischemic kidney injury in the mouse to study the trade-off using both experimental and mathematical analyses. By multi-variable molecular, morphological, and functional assessment, we identified three distinct response types of kidney tissue with respect to its adaptation to the insult: full recovery of the healthy tissue (PARS), functional recovery with residual tissue alterations (iPARS), and progressive chronic disease with fibrosis (MARS).

It is increasingly believed that the acute and chronic forms of kidney disease do not reflect distinct types of diseases with different etiologies. Instead, both forms are distinct responses to the same type of insult, such as ischemia, occurring at varying intensity. Thus, the system is a dynamical one producing qualitatively distinct behaviors in response to a quantitative continuum of perturbation. The emergence of multiple qualitatively distinct behaviors is a central characteristic of non-linear dynamical systems. Therefore, dynamical systems theory is an appropriate method for analyzing the system representing the kidney injury response.

In constructing a mathematically tractable model, we focused on the repair of the cellular damage mediated by well-characterized cell-to-cell interactions. The dynamics of the cell interaction network (Fig. 2A) exhibited multistable dynamics that mapped to the four observed disease fates: PARS, iPARS, MARS and death. Theoretical analysis of the model made a number of predictions that were experimentally confirmed and explained long-observed phenomena (Fig. 6 and Table S3). Multistability and memory are the most elementary manifestations of non-linear dynamics and require positive feedback loops in the system architecture [38]. Memory is manifest in the irreversibility, which emerges due to hysteresis in the dynamics. Physiologically, memory is represented in iPARS as the remnants of tissue pathology that persist long after the end of ischemia and in MARS as the progression to fibrosis in MARS. Memory is also represented by the protective priming effect of pre-conditioning, a phenomenon seen in several ischemic organ diseases. Thus, the kidney repair program keeps a "memory" of previous injury and is primed for a more rapid response to subsequent injury but at the expense of increased risk of fibrosis.

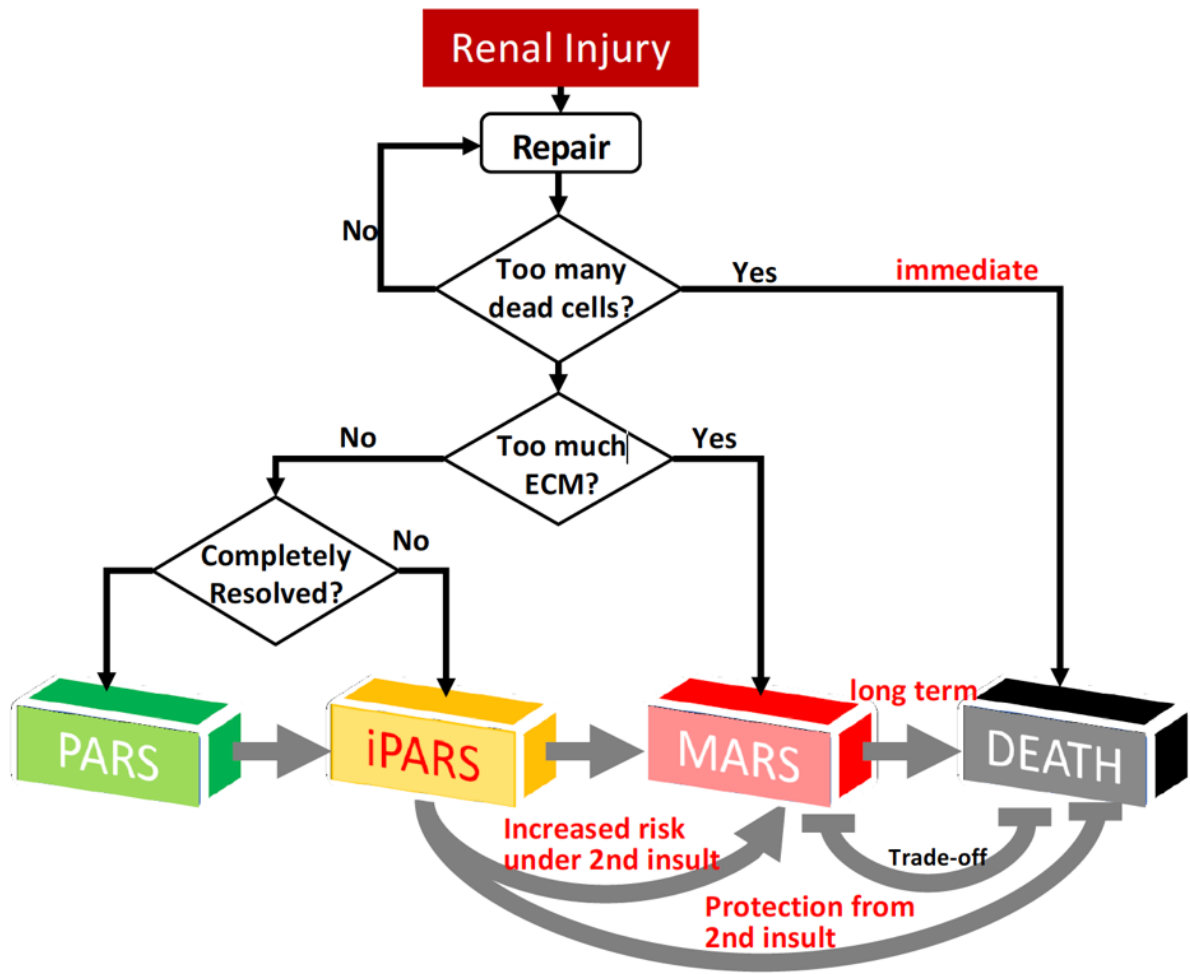

*Figure 6.* *Summary of the renal repair process. In response to injury, the renal repair system is initialized to reduce the damage through cell regeneration. If too many tubular cells are dead without timely replenishment, immediate death may happen. Otherwise, the repair process continues and generates myofibroblast that produce extracellular matrix (ECM). If too much ECM accumulates, the system goes into MARS and develops renal fibrosis. If the ECM level is low and the level of myofibroblasts does not reset to the pre-injury level, the system goes into iPARS. If all the components are completely reset to their normal states, the system goes into PARS.*

Although our model produced the observed dynamics that underlies the pathogenesis, as with all models, a simplification was necessary to enable computational study. Thus, our model does not explicitly consider many molecular mechanisms known to take part in the response. Specifically, many inflammatory cytokines, growth factors, and autacoids were not incorporated [39]. By focusing on regeneration that repairs the cellular deficit caused by ischemic destruction, we omitted inflammatory infiltrates and ensuing angiogenesis. Future studies that implement these processes and their actively controlled termination may offer opportunities to fine-tune the disease course produced by our core model, which was sufficient to robustly predict key dynamic features of the repair response. Finally, our cell-cell interaction model describes tissue-level changes (defined by cell and matrix composition) and does not possess the granularity to describe gene expression changes and associated epigenetic alterations that govern the cell state dynamics and have been implicated in the memory of AKI [40-42].

The double-edged nature of molecular players that could serve as therapeutic targets poses a dilemma for treatment. Specifically, Wnt represents an important lever point for pharmacological intervention. However, Wnt plays opposing roles, which are captured in our model: In the acute phase, it is a positive regulator of regeneration by promoting stem-like states and stimulating myofibroblast expansion. However, the latter role of Wnt signaling drives fibrosis in the chronic phase. Some studies indicate renal tubular cells undergoing pEMT are a potential target for intervention as tubular cells in pEMT state secrete various cytokines to promote the fibrosis [27,28], and others focus on the activation of resident fibroblasts as site for intervention because these cells are a source of the fibrogenic myofibroblasts [27-29,43,44]. Both mathematical modeling and experimental studies have identified key pEMT regulators, such as Snail1, HIPK2, NF-κB, and Twist1 [27,28,45-47], which may be combined with other targets, such as Wnt and Shh, for designing optimal treatment against both AKI and CKD [48]. However, these approaches commonly rest on the conventional paradigm of linear causation embodied by molecular pathways that only need to be manipulated in one direction (enhanced or inhibited), ideally at multiple points in the network by combination therapy.

However, such unidirectional up- or downregulation of pathways cannot address the complexity of a trade-off response. Because of the double-edged effect, a unidirectional (unimodal) intervention in the repair program is unlikely to achieve speedy repair and complete resolution, free of unintended consequences. Thus, we focused on the double-edged-sword effect of Wnt as a potential target, for which both agonists and antagonists exist. We formulated a treatment design as an optimization problem. Our computational optimization procedure proposed a regimen that we implemented in the animal AKI model, demonstrating that a temporal combination of applying a Wnt agonist and a Wnt antagonist thereafter can both prevent death by promoting rapid repair as well as reduce the development of fibrosis. That is, Wnt antagonists could prevent fibrosis. However, Wnt signaling is necessary to activate regenerative programs. Indeed, in studies with mice, Wnt agonists administered prior to an ischemic episode reduce kidney damage and improve renal function[36]. The key to using Wnt pathway-targeted therapies is timing the application of the agonists and antagonists. Agonists need to be administered early after the injury to promote repair; whereas antagonists need to be administered later to enable resolution and limit fibrosis.

While the broader applicability of our procedure of optimizing the dual use of antagonist and agonist to control the non-linear dynamics of the repair process remains to be further evaluated in other disease models, our results suggest that, to confront a pathogenetic mechanism that involves a target that is a double-edged sword, we need to think beyond the standard unidirectional inhibition (or activation) of a single target or pathway. As a first step, our work establishes the rationale for a new modality of non-monotonical pharmacological intervention to control non-linear disease dynamics by biphasic application of both agonists and antagonists of the same target.

**MATERIALS AND METHODS**
**Mathematical modeling and computer simulation for Renal Homeostasis and Fibrosis**
A minimal mathematical model was built to describe the renal homeostasis and fibrosis. The model considers explicitly a tubular module and a fibroblast module. In the tubular module, normal tubular epithelial cell (E), injured tubular cell (IE), and partial EMT tubular cell (PE) are considered, and the total number of epithelium sites is assumed to be a constant. In the fibroblast module, resident fibroblast, activated fibroblast, myofibroblast are considered. Communications between two modules are through secreted growth factors, Shh and Wnt. See justification of this minimal model, details of the mathematical equations and the justification of the parameters in the SUPPLEMENTARY MATERIALS. The strength of the insult for the simulation is set as 400 (in arbitrary unit, a.u.) except as indicated otherwise. To mimic the heterogeneity at the population level, the value of the parameter set is randomly and uniformly chosen from 80 to 120% of its default value with Latin Hypercubic sampling. Ward linkage is used for the hierarchical clustering analysis of the marker levels.

**The fate of organism death in the mathematical model.** Our mathematical model does not treat the complex process of organism death explicitly. Instead we used the cumulative sum of dead tubular cell (O) ( $\int_{t=0}^{t=30day} O dt$ ) as an indicator of death by considering that the death is induced by death-inducing factors leaked at the sites of other dead tubular cells. As shown in Fig. S2A, a mouse assumes a fate of death if the cumulative sum of dead tubular cell is larger than a threshold value ($\text{Cumsum}_\theta$).

**Procedure of searching the optimal dynamic treatment design for AKI by targeting on Wnt.**
For finding an optimal therapy design, we performed Metropolis search with a simulated annealing procedure in the parameter space of two drugs, Wnt and inhibitor (see details in the SUPPLEMENTARY MATERIALS). The following score function is used to quantify the effectiveness of the treatment design (see details of definition of risks in the SUPPLEMENTARY MATERIALS).

**Animal models**
Male BALB/c mice weighing about 22–25 g were obtained from the Envigo (Somerset, NJ). Renal IRI was performed in mice by using an established protocol, as described elsewhere [49]. Briefly, bilateral renal pedicles were clamped for designed timing using microaneurysm clamps to generate acute injury. During the ischemic period, body temperature was maintained at 37°C by using a temperature-controlled heating system. Animals were then administered intraperitoneally with buprenorphine at 0.05 mg/kg body wt. For Wnt pretreatment, the mice were subjected to a single intravenous injection of Wnt1 expression plasmid (pHA-Wnt1; Upstate Biotechnology) at 1 mg/kg body wt using the hydrodynamic–based gene transfer technique as reported previously [23]. For pharmacologic inhibition experiments, mice were daily intraperitoneal injection of ICG-001-phosphate (kindly provided by Dr. M. Kahn, University of Southern California, Los Angeles, CA) at 5 mg/kg body weight from the 4th day after IRI. Mice were sacrificed at 30th day after IRI, and serum and kidney tissues were collected for various analyses. Animal experiments were approved by the Institutional Animal Care and Use Committee at the University of Pittsburgh.

**Determination of Serum Creatinine**
Serum was collected from mice at different times after IRI as indicated. Serum creatinine level was determined by use of a QuantiChrom creatinine assay kit, according to the protocols specified by the manufacturer (BioAssay Systems, Hayward, CA). The level of serum creatinine was expressed as milligrams per 100 ml (dl).

**Reverse transcriptase (RT) and real-time PCR**

Total RNA isolation and quantitative, real-time RT-PCR (qRT-PCR) were carried out by the procedures described previously [20]. Briefly, the first strand cDNA synthesis was carried out by using a Reverse Transcription System kit according to the instructions of the manufacturer (Promega, Madison, WI). qRT-PCR was performed on ABI PRISM 7000 Sequence Detection System (Applied Biosystems, Foster City, CA). The PCR reaction mixture in a 25-µl volume contained 12.5 µl 2x SYBR Green PCR Master Mix (Applied Biosystems), 5 µl diluted RT product (1:10) and 0.5 µM sense and antisense primer sets. PCR reaction was run by using standard conditions. After sequential incubations at 50°C for 2 min and 95°C for 10 min, respectively, the amplification protocol consisted of 40 cycles of denaturing at 95°C for 15 sec, annealing and extension at 60°C for 60 sec. The standard curve was made from series dilutions of template cDNA. The mRNA levels of various genes were calculated after normalizing with β-actin. Primer sequences used for amplifications were presented in Table S4.

**Histology and immunohistochemical staining**

Paraffin-embedded mouse kidney sections (3-µm thickness) were prepared by a routine procedure. The sections were stained with Periodic acid–Schiff (PAS) staining reagents by standard protocol. Immunohistochemical staining was performed according to the established protocol as described previously [50]. The antibodies against Wnt1 (ab15251); α-SMA (ab5694) (Abcam, Cambridge, MA), FSP-1 (#07-2274; EMD Millipore, Burlington, MA), Vimentin (#5741); PDGFR-β (#3169) (Cell Signaling Technology, Danvers, MA) were used.

**Statistical analyses**

All data were expressed as mean ± SEM. Statistical analysis of the data was performed using SigmaStat software (Jandel Scientific Software, San Rafael, CA). Comparison between groups was made using one-way ANOVA, followed by the Student-Newman-Keuls test. $P < 0.05$ was considered significant.

**ACKNOWLEDGMENTS:** We thank Nancy R. Gough (BioSerendipity, LLC) for critical discussions and editorial assistance.

**Funding:** The project was supported by the National Science Foundation through DMS-1462049, the National Institutes of Health through Grant Number UL1TR001857, DK064005, DK106049, and DK119232, the National Science Foundation of China grant 81521003, and the Guangdong Science Foundation Innovative Group Grant 2014A030312014. H.F. was supported by the National Science Foundation of China grant 81770737.

**AUTHOR CONTRIBUTIONS:**

Conceptualization, X.-J.T., D.Z., Y. L., and J. X.; Methodology, X.-J.T., D.Z., H.F., R. Z., and X. W.; Investigation, X.-J.T., D.Z., H.F., R.Z., X.W., Y. L., and J. X.; Writing – Original Draft, X.-J.T., D.Z. and J.X.; Writing – Review & Editing, X.-J.T., D.Z. Y.L., J.X. and S.H.; Funding Acquisition, Y.L. and J.X.; Resources, Y.L. and J.X.; Supervision, Y.L. and J.X.

**DECLARATION OF INTERESTS**

The authors declare no conflict of interest.